\documentclass{ws-procs975x65}

\begin{document}

\title{Microlensing towards M31:\\ candidates and perspectives}

\author{S. Calchi Novati\footnote{\uppercase{W}ork 
supported by the \uppercase{S}wiss 
\uppercase{N}ational \uppercase{S}cience \uppercase{F}oundation
and by the \uppercase{T}omalla \uppercase{F}oundation.}
 \\ \footnotesize{(on behalf of the SLOTT-AGAPE and the POINT-AGAPE collaborations)}}

\address{Institute of Theoretical Physics, University of Z{\"u}rich, Winterthurestrasse 190, 8057 Z{\"u}rich, Switzerland\\
E-mail: novati@physik.unizh.ch}

\maketitle

\abstracts{
Recent results of the SLOTT-AGAPE and POINT-AGAPE collaborations
on a search for microlensing events in direction of the Andromeda galaxy,
by using the pixel method, are reported.  The detection of 4 microlensing events, 
some likely to be due to self--lensing, is discussed. One microlensing light curve is shown to be 
compatible with a binary lens. The present analysis still
does not allow us to draw conclusions on the MACHO content of the
M31 galaxy.}

\section{Introduction}

Since Paczy\'nski's original proposal\cite{pacz}
gravitational microlensing has been proben to be
a powerful tool for the detection of the dark matter
component in galactic haloes in the form of MACHOs.
Searches in our Galaxy towards LMC\cite{macho,eros} show
that up to 20\% of the halo could be formed
by objects of around $M \sim 0.4\,M_\odot$,
but these results are still debated\cite{jetzer}.

Searches towards M31, nearby and similar to our Galaxy,
have also been proposed\cite{crotts92,agape93,jetzer94}.
This allows to probe a different line of sight in our Galaxy,
to globally test M31 halo and, furthermore,
the high inclination of the M31 disk is expected
to provide a strong signature (spatial distribution) for halo microlensing signals.

Along a different direction, results of a microlensing survey towards M87, 
where one can probe both the M87 and the Virgo cluster haloes, 
have also been presented\cite{m87}.

For extragalactic targets, due to the distance, the sources for microlensing signals
are not resolved. This claims for an original technique, the \emph{pixel method}, 
the detection of flux variations of unresolved sources\cite{agape97,agape99,tom96},
the main point being that one follows flux variations of every pixel
in the image instead of single stars. 

I review here the results from two different survey of M31 aimed at the detection
of microlensing events, carried out by the  SLOTT-AGAPE\cite{mdm1,mdm2}
and by the POINT-AGAPE collaborations\cite{point01,point03}.
The WeCapp\cite{wecapp,wecapp03} and the MEGA\cite{mega} collaborations have also
presented a handful of microlensing events.

\section{Pixel lensing with MDM data}

The SLOTT-AGAPE collaboration has been using data collected on the 1.3m 
McGraw-Hill Telescope at the MDM observatory, Kitt Peak (USA). Two
fields, $17'\times 17'$ wide each, on the opposite side (and including) the bulge are observed
(centered in $\alpha=$ 00h 43m 24s, $\delta  = 41^{\!\circ}12'10''$
(J2000) ``Target'', on the far side of M31, and $\alpha=$ 00h 42m 14s,
$\delta  =41^{\!\circ}24'20''$ (J2000) ``Control''). Two filters, 
similar to standard $R$ and $I$ Cousins, have been used in order to test achromaticity. 
Furthermore, this particular colour information gives 
the chance of having a better check on red
variable stars, which can contaminate the search
for microlensing events. Observations have been carried out in a two years
campaign, from October 1998 to the end of December 1999.
Around 40 (20) nights of observations are available
in the Target and Control field respectively. 

To cope with photometric and seeing variations
we follow the ``superpixel photometry''\cite{agape97,mdm1} approach,
where one statistically calibrate the flux of each image
with respect of a chosen reference image.
In particular, the seeing correction is based
on an empirical linear correction of the flux,
and we do not need to evaluate the PSF of the image.

The search for microlensing events is carried out in two steps.
Through a statistical analysis on the light curve
significant flux variations above the baseline are detected,
then we perform a shape analysis on the selected light curve,
$\sim 10^3$, to distinguish between microlensing and other
variable stars. 

The background of variable sources is a main problem
for pixel lensing searches of microlensing signals. 
First, the class of stars to which we are in principle most sensitive are the red giants, 
for which a large fraction are variable stars 
(regular or irregular). Second, as looking for \emph{pixel}
flux variations, it is always possible to collect (in the same pixel)
light from more than one source whose flux is varying.
Thus, in the analysis, one is faced with two problems: 
large-amplitude variable sources whose signal can mimic a microlensing signal,
and variable sources of smaller amplitude whose signal can 
give rise to non-gaussian fluctuations
superimposed on the background or on other physical variations.

In a first analysis\cite{mdm1} we followed a conservative
approach to reduce the impact of these problems. Severe criteria
in the shape analysis with respect to the Paczy\'nski fit were adopted
(with a stringent cut for the $\chi^2$)
and, furthermore, candidates  with both a long timescale
($t_{1/2}>40$ days) and a red colour ($(R-I)_C>1$) were excluded, since these most likely
originate from variable stars.

\begin{figure}[ht]
\centerline{\epsfxsize=2.65in
\epsfbox{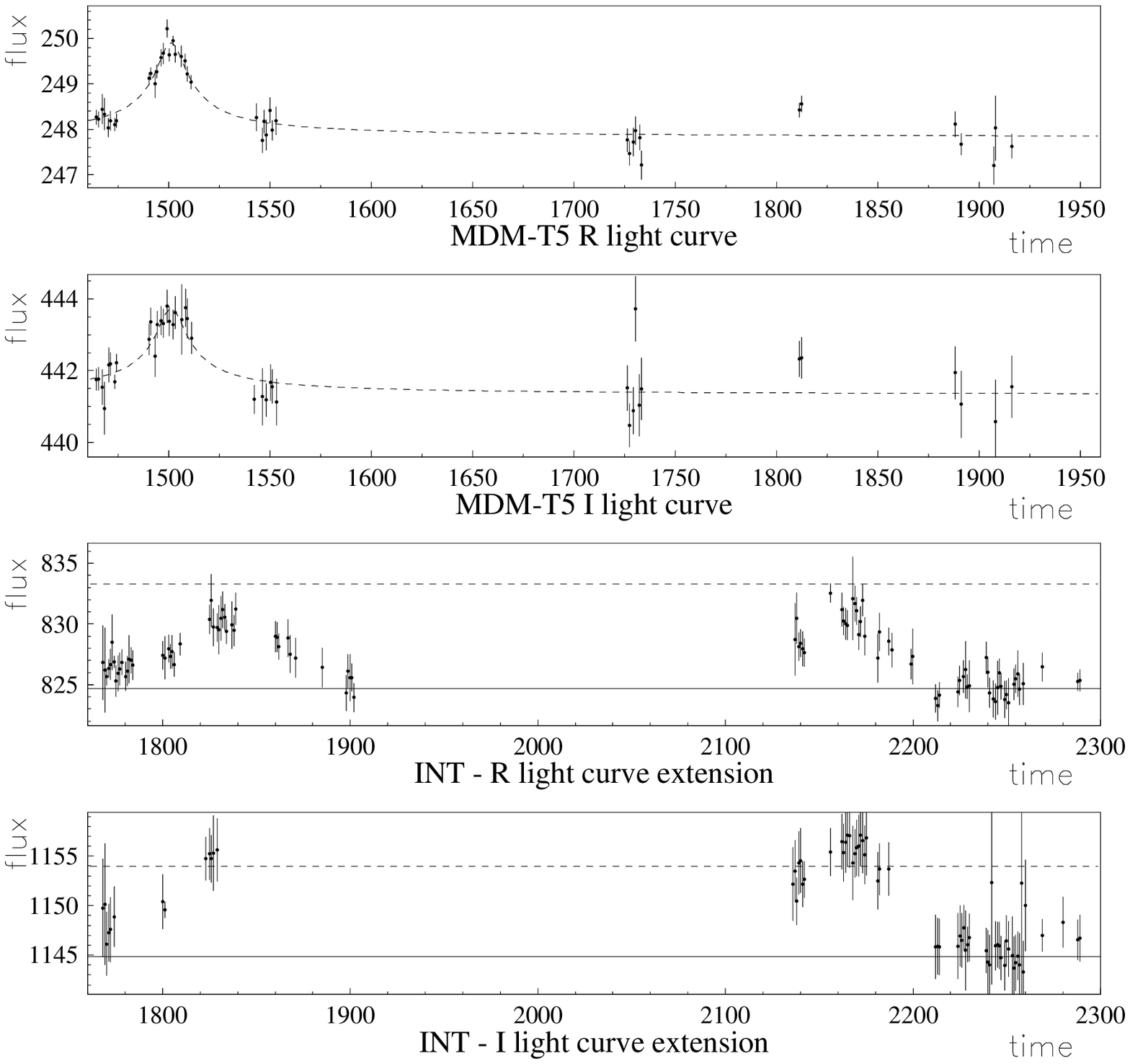}
\epsfxsize=2.5in
\epsfbox{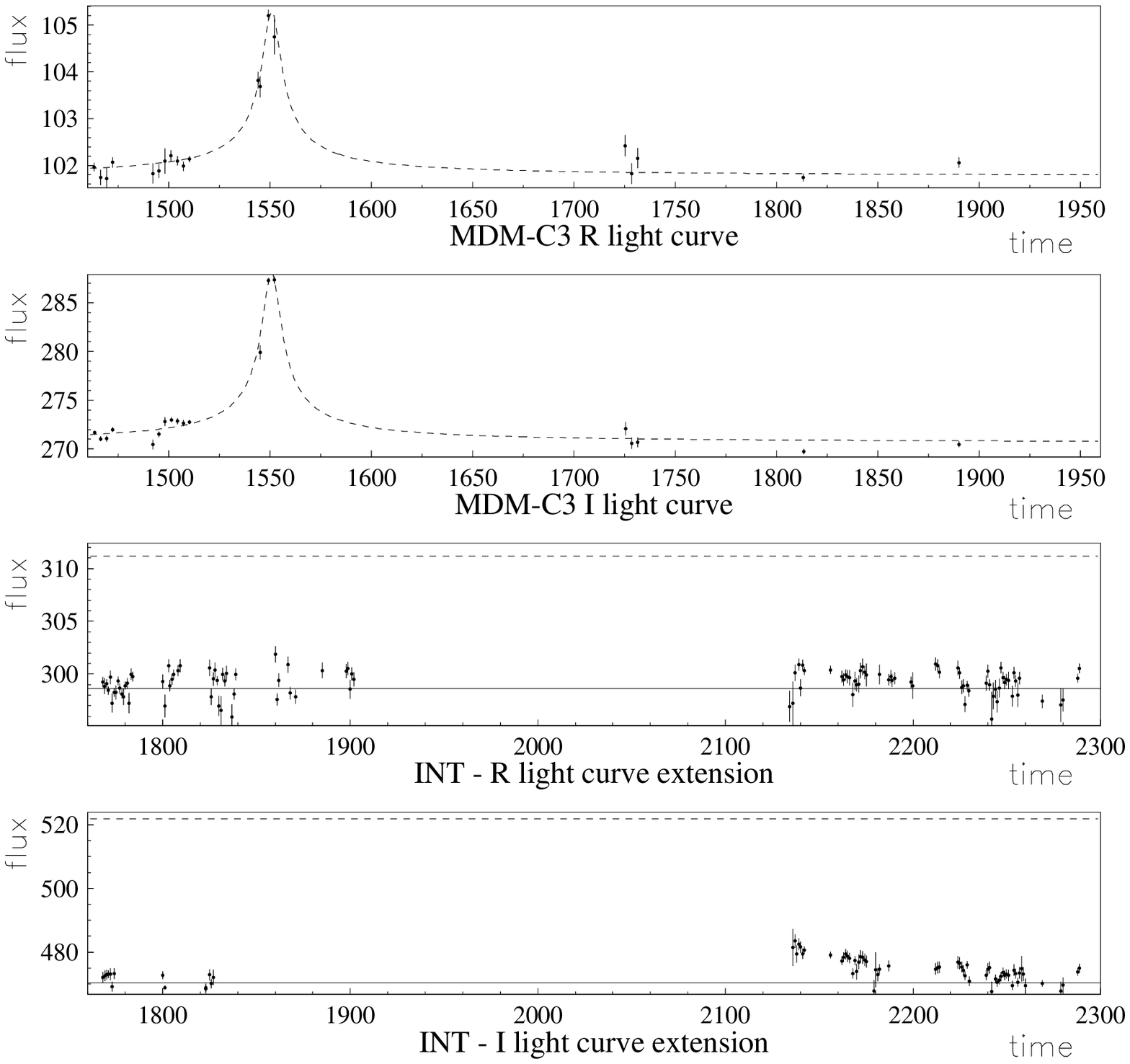}}   
\caption{MDM T5 and C3 light curves together with their extension into the INT data.
On the $y$ axis, flux is in ADU/s; on the $x$ axis, time is in days, with the origin in J-2449624.5 (both data sets). 
For the MDM light curves the dashed line represent the result of the Paczy\'nski fit.
For the INT light curves, shown together with the solid line representing
the baseline is a dashed line representing the level of the maximum deviation of flux found on the corresponding MDM light curve.}
\label{mdmcl}
\end{figure}

In this way 10 variations compatible
with a microlensing (time width in the range 15-70 days,
and flux deviation at maximum all above $\Delta R\sim 21.5$) 
were selected. However, due to the rather poor sampling and the short baseline,
the uniqueness bump requirement could not be proben efficiently.
A successive analysis\cite{mdm2} on the INT extension of these
light curves then shows that all these variations are indeed
due to variable sources and rejected as microlensing candidates. 
Indeed, in the same position, a variation with compatible time width and flux deviation
is always found on INT data. In Fig. \ref{mdmcl} (left) we show
one MDM flux variation (T5) from this selection, nicely fitting
a Paczy\'nski light curve, then its extension on the INT data 
where it is clearly seen that the bump does repeat with the same shape,
showing that this is actually a variable source.

A second analysis is then carried out where we relax
the criteria introduced to characterize the shape, as this
has proben not to efficiently reject variable stars and 
indeed could introduce a bias against real microlensing
events whose light curve might be disturbed by some
non gaussian noise, and, on the other hand, we restrict the allowed space of physical
parameters, in particular we consider only relatively short
(time width less than 20 days) flux variations (this range of
parameter space being consistent with what expected 
on the basis on Monte Carlo simulations\cite{mdm1}).

As an outcome, out of further 8 detected flux variations,
INT vetting allows to firmly exclude 5 as microlensing,  
leaving 2 light curves for which this test is considered inconclusive
and 1 lying in a region of space not covered by the INT field
(with $t_{1/2}\in (13,20)$ days and $\Delta R_{max} \in (21.0,21.8)$).

By ``inconclusive'' it is meant that a flux variation
is detected at the same position on INT data,
but where the comparison of the time width and the flux deviation
added to the rather poor sampling along the bump 
do not allow to conclude sharply on the uniqueness test,
leaving open the possibility of the detection
of a microlensing light curve superimposed on (the light curve of)
a variable star (Fig. \ref{mdmcl}, right).

\section{Microlensing events with INT data}

The POINT-AGAPE collaboration\cite{point01} is carrying out a survey of M31
by using the Wide Field Camera (WFC) on the 2.5 m INT telescope.
Two fiels, each of $\sim 0.3$ deg$^2$ are observed. The observations
are made in three bands close to Sloan $g',\,r',\,i'$. We
report here on the results from the analysis of 143
nights collected in two years between August 1999 and January 2001.
As described for MDM data, superpixel photometry is performed
to bring all the images to the same reference one,
then a similar analysis for the search of microlensing
candidates is carried out.

\begin{figure}[ht]
\centerline{\epsfxsize=4.0in
\epsfbox{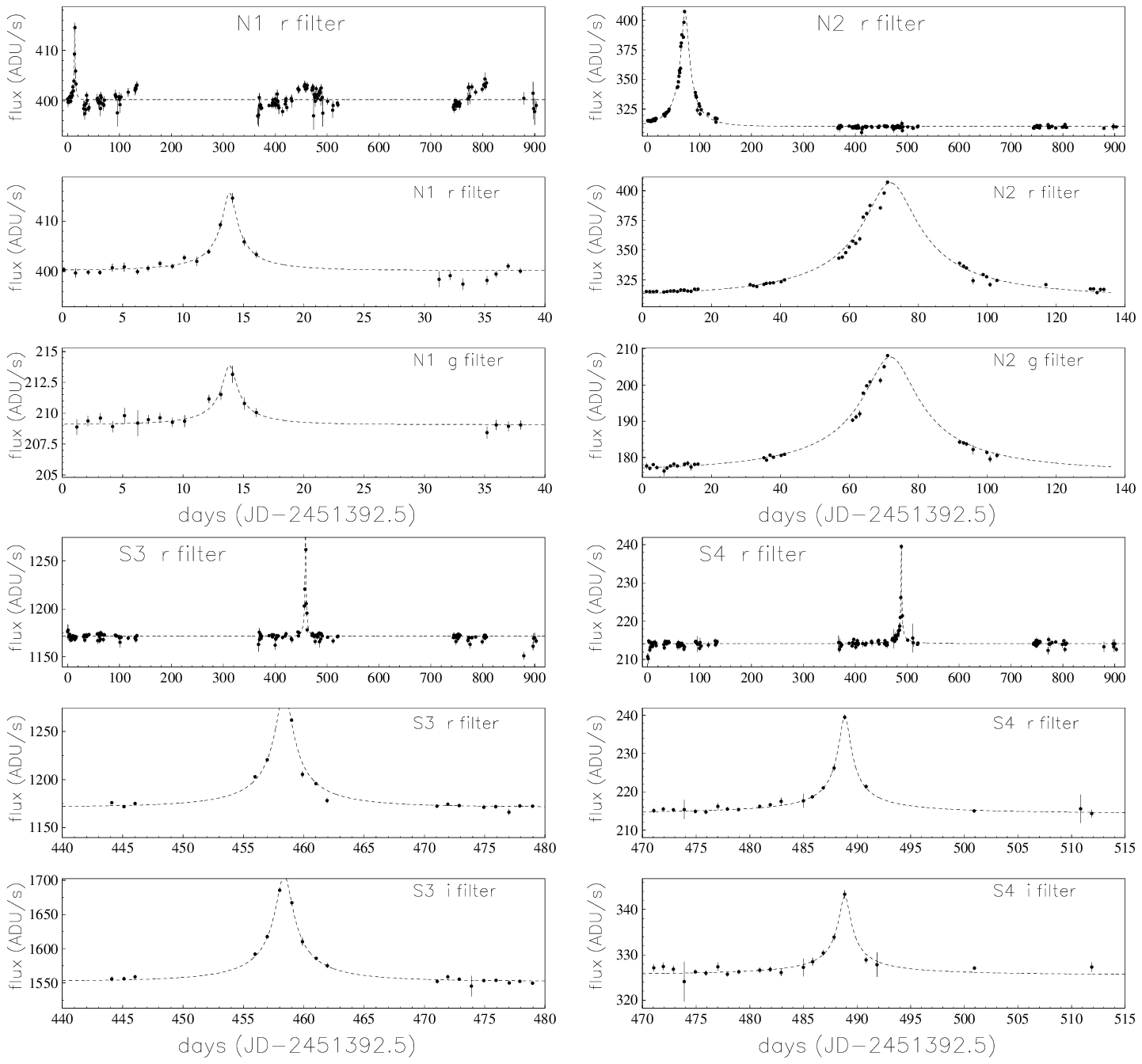}}
\caption{Three years data light curves for the 4 POINT-AGAPE microlensing events.}
\label{int4}
\end{figure}

A first analysis\cite{point03} is made with the aim
to detect short ($t_{1/2} < 25$ days) and bright variations
($\Delta R < 21$ at maximum amplification),
compatible with a Paczy\'nski signal.
The first requirement is suggested
by the results on the predicted characteristics
of microlensing events of a Monte Carlo simulation
of the experiment. As an outcome,
four light curves are detected, whose characteristics
are summarised in Table \ref{tab}, and whose light curve
are shown in Fig. \ref{int4} (with a third year data
added). We stress that their
signal is incompatible with any known variable star,
therefore it is safe to consider these as viable
microlensing events.

\begin{table}[h]
\tbl{Characteristics of the four microlensing events
detected by the POINT-AGAPE collaborations. $d$ is the projected
distance from the center of M31.\label{tab}}
{\begin{tabular}{@{}ccccc@{}} \toprule
 & PA-99-N1 & PA-99-N2 & PA-00-S3 &
PA-00-S4 \\ \colrule
$\alpha$ (J2000) & 00h42m51.4s & 00h44m20.8s & 00h42m30.5s & 00h42m30.0s \\
$\delta$  (J2000) & $41^\circ\, 23'\, 54''$ & $41^\circ\, 28'\, 45''$ &
           $41^\circ\, 13'\, 05''$ & $40^\circ\, 53'\, 47''$\\
$d$      & $7'\, 52''$ & $22'\, 03''$ & $4'\, 00''$ & $22'\, 31''$\\
$t_{1/2}$ (days) & $1.8\pm 0.2$ & $21.8\pm 0.2$ & $2.2\pm 0.1$ & $2.1\pm 0.1$  \\
$\Delta R_{max}$ & $20.8\pm 0.1$ & $19.0\pm 0.2$ & $18.8\pm 0.2$ & $20.7\pm 0.2$  \\
\botrule
\end{tabular}}
\end{table}

Once a microlensing event is detected it is important,
given the aim to probe the halo content in form
of MACHO, to find out its origin, namely, whether it is due to
self-lensing within M31 or to a MACHO. This is not
straightforward. The spatial distribution of the events
is an important tool, but still unusable given the small statistic.
The observed characteristics of the variations to some extent
can give a hint on the nature of the lens, but again,
the small number of detected events so far makes this 
approach rather unviable. However, we stress that the detection 
of some self-lensing event, as they are expected to be found (their existence
being predicted only on the basis of the rather well known
luminous component of M31), is essential to assess
the efficiency of the analysis. In the following,
starting from their spatial position, we briefly
comment on each of the detected events.

PA-99-N1: For this event it has been possible
to identify the source on HST archival images. 
The knowledge of the flux of the unamplified source allows
to break the degeneracy between the Einstein
crossing time and the impact parameter
for which one obtains the values $t_E=9.7 \pm 0.7$
and $u_0 = 0.057 \pm 0.004$. The baseline
shows two secondary bumps: they are due to
a variable star lying some 3 pixels away from the 
microlensing variation. The position of this event,
$7'\,52''$ from the center of M31, makes
unlikely the hypothesis of bulge-bulge self-lensing.
The lens can be either a MACHO (with equal chance
in the M31 or the Milky Way halo) or a low mass
($\sim 0.2\,M_\odot$) disk star with the source
lying in the bulge. The first  case is more likely 
assuming a halo fraction in form of MACHOs above 20\%.

PA-99-N2: This variation lies at some $22''$
away from the center of M31, therefore
is an excellent microlensing MACHO  candidate.
However, this variation turns out to be
almost equally likely to be due to
disk--disk self lensing. This light curve is particularly
interesting because it shows clear deviations
from a Paczy\'nski shape, while remaining
achromatic (and unique) as expected for a microlensing event.
We recall that this shape is characteristic
for variations where the point-like (source and lens)
and uniform motion hypothesis hold. After exploring\cite{point03a}
different explanations, it is found that the observations
are consistent with an unresolved RGB or AGB star in M31
being microlensed by a binary lens, with
a mass ratio of $\sim 1.2\times 10^{-2}$. An analysis
of the relative optical depth shows that a halo
lens (whose mass is estimated to lie in the range
0.009-32 $M_\odot$) is more likely
than a stellar lens (with mass in this case expected
in the range 0.02-3.6 $M_\odot$) provided that 
the halo mass fraction in form of compact objects
is at least around $15\%$.

PA-00-S3: This event is the nearest found so far
from the center of M31 ($d=4'\,00''$). Its extension
past in time on MDM data shows no variations. The good sampling
along the bump allows to get a rather robust
estimation of the Einstein time, $t_E=13\pm 4$ days.
This value, together with its position, makes
the bulge-bulge self-lensing hypothesis
the most likely for this event. 

PA-00-S4: This event is found far away from the M31 center,
but only at $2'\,54''$ from the center of the dwarf galaxy
M32. A detailed analysis\cite{point02} shows that the source
is likely to be a M31 disk A star, the main evidence being
the observed rather blue colour $(R-I) = 0.0\pm 0.1$. Given that
M32 lies $\sim$ 20 kpc in front of M31, the study of the relative
optical depth allows to conclude that the most likely
position for the lens is M32. 

\section{Conclusions}

As a general outcome of the results presented, we stress that
the detection of microlensing events towards M31 is now established.
The open issue to be still explored is the 
study of the M31 halo fraction in form of MACHOs.
With respect to this analysis, the events detected so far are all compatible with
stellar lenses, but the MACHO hypothesis is still open,
and we recall that the analysis for the INT data is still not concluded
(besides a third year data, variations with $\Delta R_{max}>21$ 
have still to be studied). Once this analysis completed, it is the necessary to ``weight''
it with an efficiency study of the pipeline of detection 
before meaningfully comparing its results with the prediction
of a Monte Carlo simulation. This should eventually
allow us to draw firm conclusions
on the halo content in form of MACHO of M31.

\end{document}